\documentstyle[twoside,epsfig]{elsart}
 
\journal{Astroparticle Physics}
\begin{document}
\begin{frontmatter}    
\title{Conceptual Design of a Next-Generation All-Sky 
$\gamma$-ray Telescope Operating at TeV Energies}
\author[losalamos]{R.S. Miller\thanksref{email}},
\author[santacruz]{S. Westerhoff}
\address[losalamos]{Neutron Science \& Technology, 
Los Alamos National Laboratory, Los Alamos, NM\,87545, USA} 
\address[santacruz]{Santa Cruz Institute for Particle Physics, 
University of California, Santa Cruz, CA\,95064, USA}

\thanks[email]{corresponding author: richard@lanl.gov}
 
\begin{abstract}
The next generation all-sky monitor operating at TeV 
(0.1\,-30\,TeV) 
energies should be capable of performing a continuous 
high sensitivity 
sky survey, and detecting transient sources, such as 
AGN flares, with high statistical significance
on timescales of hours to days. 
We describe an instrument concept for a large area,
wide aperture, TeV ground-based $\gamma$-ray telescope.  
The conceptual design is comprised of 
$\sim$\,20\,000 $1\,\mathrm{m}^{2}$ 
scintillator-based pixels arranged in a densely packed mosaic. 
In addition to fast timing, the dense array of pixels also 
provides a unique imaging capability; the shower image, 
the spatial distribution of secondary particles at ground level, 
can be exploited to identify and reject hadronic backgrounds. 
The good angular resolution and background 
rejection capabilities lead directly to high
sensitivity, while retaining a large field of view, 
large effective 
area, and high duty cycle. An instrument such as the one described 
here complements the narrow field of view air-Cherenkov telescopes 
and could, in conjunction with future
space-based instruments, extend the energy range for continuous 
monitoring of the $\gamma$-ray sky from MeV through TeV energies.
\end{abstract}
\begin{keyword} $\gamma$-ray telescopes; cosmic-ray detectors; 
extensive air showers
\end{keyword}
\end{frontmatter}

\newpage

\section{Introduction}

Very High Energy (VHE) $\gamma$-ray astronomy \cite{ong} is still 
in its infancy. Operating in the energy 
range from $\sim$\,100 GeV to 30 TeV and beyond, this subfield of 
astronomy represents an exciting, 
relatively unsampled region of the electromagnetic spectrum, and a 
tremendous challenge. To develop more 
fully this field needs an instrument capable of performing 
continuous systematic sky surveys and 
detecting transient sources on short timescales without 
a priori knowledge of their location. These primary
science goals require a telescope with a wide field of view 
and high duty cycle, excellent
source location and background rejection capabilities - 
an instrument that complements both existing and future ground- 
and space-based $\gamma$-ray telescopes.

To be viable VHE astronomy must overcome a number of fundamental 
difficulties. Since the flux of VHE photons is small, telescopes 
with large collecting areas ($\>10^3\,\mathrm{m}^2$)
are required to obtain statistically significant photon samples; 
telescopes of this size can, currently, 
only be located on the Earth's surface. However, 
VHE photons do not readily penetrate the $\sim\,28$ 
radiation lengths of the Earth's atmosphere 
($1030\,\mathrm{g}/\mathrm{cm}^2$ thick at sea-level) but 
instead interact with air molecules to produce secondary 
particle cascades, or extensive air showers. 
Another difficulty of VHE astronomy is the large background 
of hadronic air showers, induced by 
cosmic-ray primaries (primarily protons), that cannot be vetoed.

In this paper we describe the conceptual design of an instrument 
that builds upon traditional extensive air shower methods; 
however, unlike typical extensive air shower arrays the detector 
design utilizes unique imaging capabilities 
and fast timing to identify (and reject) hadronic cosmic-ray 
backgrounds and achieve excellent angular 
resolution, both of which lead to improved sensitivity.
In the following sections we briefly motivate the need for such 
an instrument 
(Section 2), discuss in detail telescope design parameters with 
emphasis on their optimization (Section 3), 
describe the conceptual design of a VHE telescope and the 
simulations used in this study (Section 4), and 
evaluate the capabilities of such a detector in terms of source 
sensitivity (Section 5). Finally, the results 
of this study are summarized and compared to both current and 
future VHE telescopes.

\section{Motivation}

VHE $\gamma$-ray astronomy has evolved dramatically in the 
last decade with the initial detections of steady 
and transient sources, galactic and extragalactic sources. 
To date 7 VHE $\gamma$-ray sources have been 
unambiguously detected 
\cite{crab,mrk421,mrk501,1ES2344,p1706,vela,sn1006}; 
this contrasts dramatically with 
the number of sources detected in the more traditional 
regime of $\gamma$-ray astronomy at energies below 
$\sim$\,20 GeV. The EGRET instrument aboard the 
Compton Gamma-Ray Observatory, for example, has detected 
pulsars, supernova remnants, gamma-ray bursts, active 
galactic nuclei (AGN), and approximately 50 unidentified 
sources in the 100 MeV-20 GeV range \cite{agn1,agn2}. 
The power-law spectra of many EGRET sources show no sign 
of an energy cutoff, suggesting that they may be observable 
at VHE energies. 

The 4 Galactic VHE objects, all supernova remnants, appear 
to have $\gamma$-ray emission that is constant in both 
intensity and spectrum. The 3 extragalactic VHE sources are 
AGN of the blazar class. Although AGN have been 
detected during both quiescent and flaring states, it is the 
latter that produce the most statistically 
significant detections. During these flaring states the 
VHE $\gamma$-ray flux has been 
observed to be as much as 10 times that of the Crab nebula - 
the standard candle of TeV astronomy \cite{hegra_flare}.

Although observed seasonally since their initial detection, 
long term continuous monitoring of the TeV sources 
detected to date has never been possible, nor has there ever 
been a systematic survey of the VHE sky. This is 
primarily due to the fact that all VHE source detections to 
date have been obtained with air-Cherenkov telescopes. 
Because they are optical instruments, air-Cherenkov telescopes 
only operate on dark, clear, moonless nights - 
a $\sim\,5-10\,\%$ duty cycle for observations; these 
telescopes also have relatively narrow fields of view 
($\sim\,10^{-2}\,\mathrm{sr}$). Although they are likely 
to remain unsurpassed in sensitivity for detailed 
source observations these telescopes have limited usefulness 
as transient monitors and would require over a 
century to complete a systematic sky survey. 

The identification of additional VHE sources would contribute 
to our understanding of a range of unsolved 
astrophysical problems such as the origin of cosmic rays, 
the cosmological infrared background, and the nature 
of supermassive black holes. Unfortunately, the field of VHE 
astronomy is data starved; new instruments 
capable of providing continuous observations and all-sky 
monitoring with a sensitivity approaching that
of the air-Cherenkov telescopes are therefore required. A VHE 
telescope with a wide field of view and high duty 
cycle could also serve as a high-energy early warning system, 
notifying space- and ground-based instruments of 
transient detections quickly for detailed multi-wavelength 
follow-up observations. Its operation should coincide 
with the launch of the next-generation space-based instrument 
such as GLAST \cite{glast}. 

\section{Figure of Merit Parameters}

A conceptualized figure of merit is used to identify the 
relevant telescope design parameters. This 
figure of merit, also called the signal to noise ratio, 
can be written as
\begin{equation}
\left(\frac{signal}{noise}\right) \propto 
\frac{R_\gamma Q \sqrt{A^{eff}~T}}{\sigma_\theta} 
\label{equation1}
\end{equation}

\begin{table}
\caption{\label{table1} Figure of merit parameter definitions.}
\vskip 0.5cm
\small
\centerline{
\begin{tabular}{|l l l|} 
\hline
{\em Parameter} & {\em Units}    & {\em Definition} \\ 
\hline
\hline
$A_{eff}$       & $\mathrm{m}^2$ & (effective) detector area \\ 
$T$             & sec            & exposure \\ 
$\sigma_\theta$ & $^o$           & angular resolution \\ 
$R_\gamma$      & -              & $\gamma$/hadron 
relative trigger efficiency \\ 
$Q$             & -              & $\gamma$/hadron 
identification efficiency \\ 
\hline
\end{tabular}}
\end{table}
\normalsize
where the various parameters are defined in 
Table\,\ref{table1}. Ultimately, source sensitivity is 
the combination of these design parameters. 
Although a more quantitative 
form of the figure of merit is used to estimate the 
performance of the conceptual telescope design (see 
Equation\,\ref{equation4}), we use Equation\,\ref{equation1} 
to address specific design requirements. 

\subsection{$R_\gamma$}
\label{sec_alt}

Air showers induced by primary particles in the 
100\,GeV to 10\,TeV range 
reach their maximum particle\footnote{Throughout the rest 
of the paper the generic term ``particles'' will
refer to $\gamma, e^{\pm}, \mu^{\pm}$, and hadrons unless 
otherwise noted.} number typically at altitudes 
between 10 and 15\,km above sea level (a.s.l.).
An earth-bound detector therefore samples the cascade at 
a very late stage of its
development, when the number of shower particles 
has already dropped by an
order of magnitude from its value at shower maximum. 

\begin{figure}
\epsfig{file=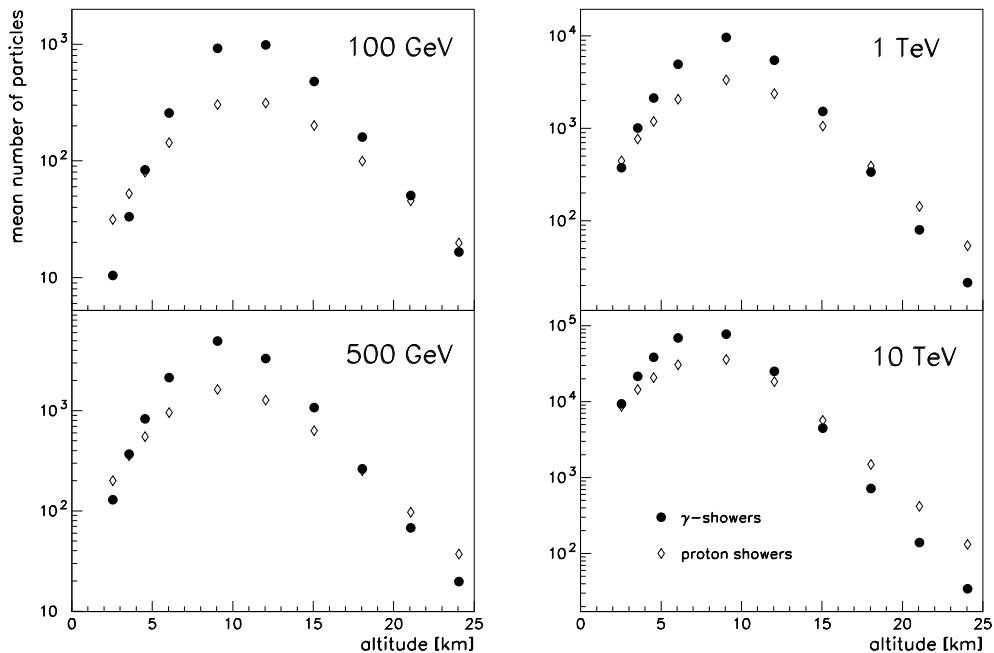,width=14.0cm}
\caption{\label{longi}
Mean number of particles 
($\gamma, e^{\pm}, \mu^{\pm}$, hadrons) vs. altitude for proton-
and $\gamma$-induced air showers with primary
energies 100\,GeV, 500\,GeV, 1\,TeV, and 10\,TeV.
The low energy cutoff of the particle kinetic energy is
100\,keV ($\gamma, e^{\pm}$), 0.1\,GeV ($\mu^{\pm}$), 
and 0.3\,GeV (hadrons).}
\end{figure}

\begin{figure}
\epsfig{file=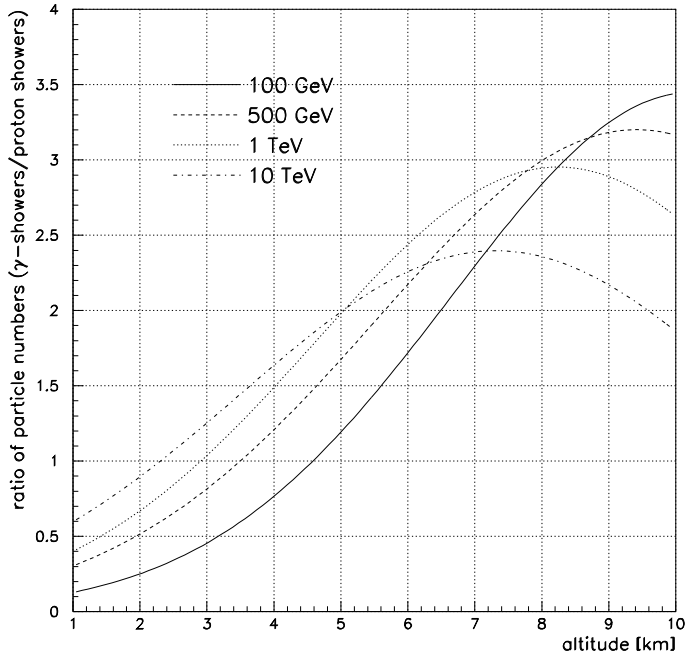,width=14.0cm}
\caption{\label{r_gamma}
Ratio of particle numbers in $\gamma$- and proton-induced 
showers vs. altitude.}
\end{figure}

Figure\,\ref{longi} shows the result of computer 
simulations\footnote {Here and in the following analysis, 
the CORSIKA 5.61 \cite{corsika} code is used for air-shower 
simulation in the atmosphere. It is briefly 
described in the next section.} of the longitudinal profile 
of air showers induced in the Earth's 
atmosphere by proton and $\gamma$-primaries with fixed 
energies and zenith angles 
$0^{\mathrm o}\le\theta\le45^{\mathrm o}$. The small number 
of particles reaching 2500m detector altitude 
sets severe limits for observations at these altitudes. 
In addition, the number of particles in proton 
showers actually exceeds the number of particles in 
$\gamma$-showers at low altitudes (Figure\,\ref{r_gamma}).
This implies that the trigger probability, and thus the 
effective area, of the detector is larger for 
proton than for $\gamma$-showers, an unfavorable situation 
which leads to an $R_{\gamma}$ (the ratio of 
$\gamma$-ray to proton trigger efficiency) less than 1. 
At 4500\,m, however, the mean number of particles 
exceeds the number at 2500\,m by almost an order of magnitude 
at all energies. Therefore, telescope location 
at an altitude $\geq$\,4000\,m is important for an air 
shower array operating at VHE energies not only 
because of the larger number of particles, and hence the 
lower energy threshold, but also because of the 
intrinsic $\gamma$/hadron-separation available, due to 
the relative trigger probabilities, that exists at higher 
altitudes. 

\begin{figure}
\epsfig{file=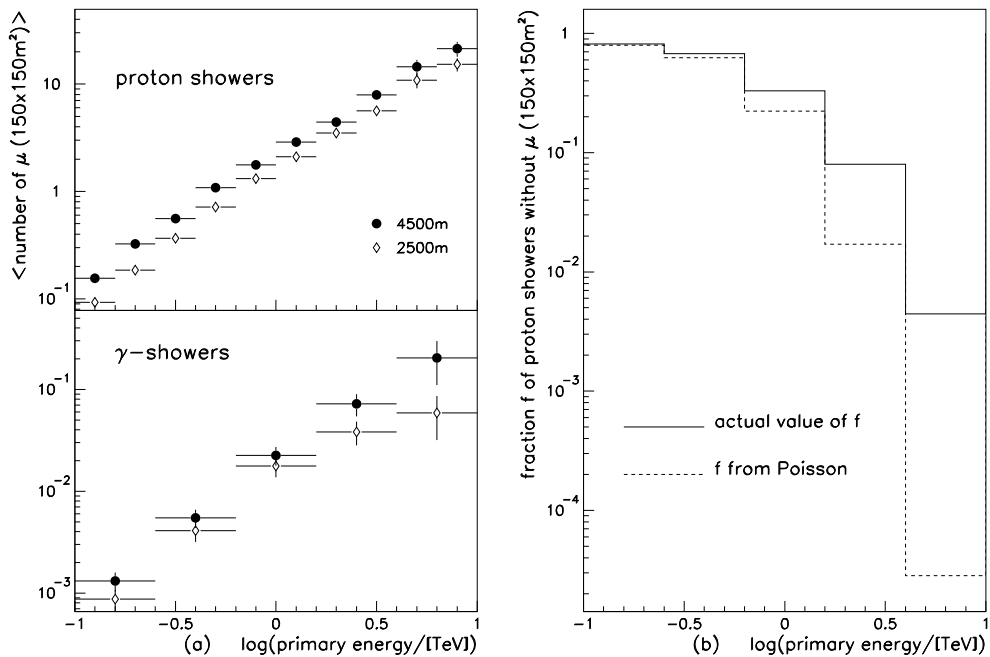,width=14.0cm}
\caption{\label{n_muon}
For a $150\times 150\,\mathrm{m}^2$ detector area and 
cores randomly distributed over the detector area,
(a) shows the mean number of $\mu^{\pm}$ in proton-induced showers
as a function of the energy of the primary particle, 
and (b) shows the fraction $f$ of proton showers
without $\mu^{\pm}$ as a function of the energy of the primary 
particle. The solid line is the actual value
of $f$, the dashed line is the expected value assuming the 
number of $\mu^{\pm}$ follows a
Poisson distribution.}
\end{figure}

\subsection{$Q$}

\begin{figure}
\epsfig{file=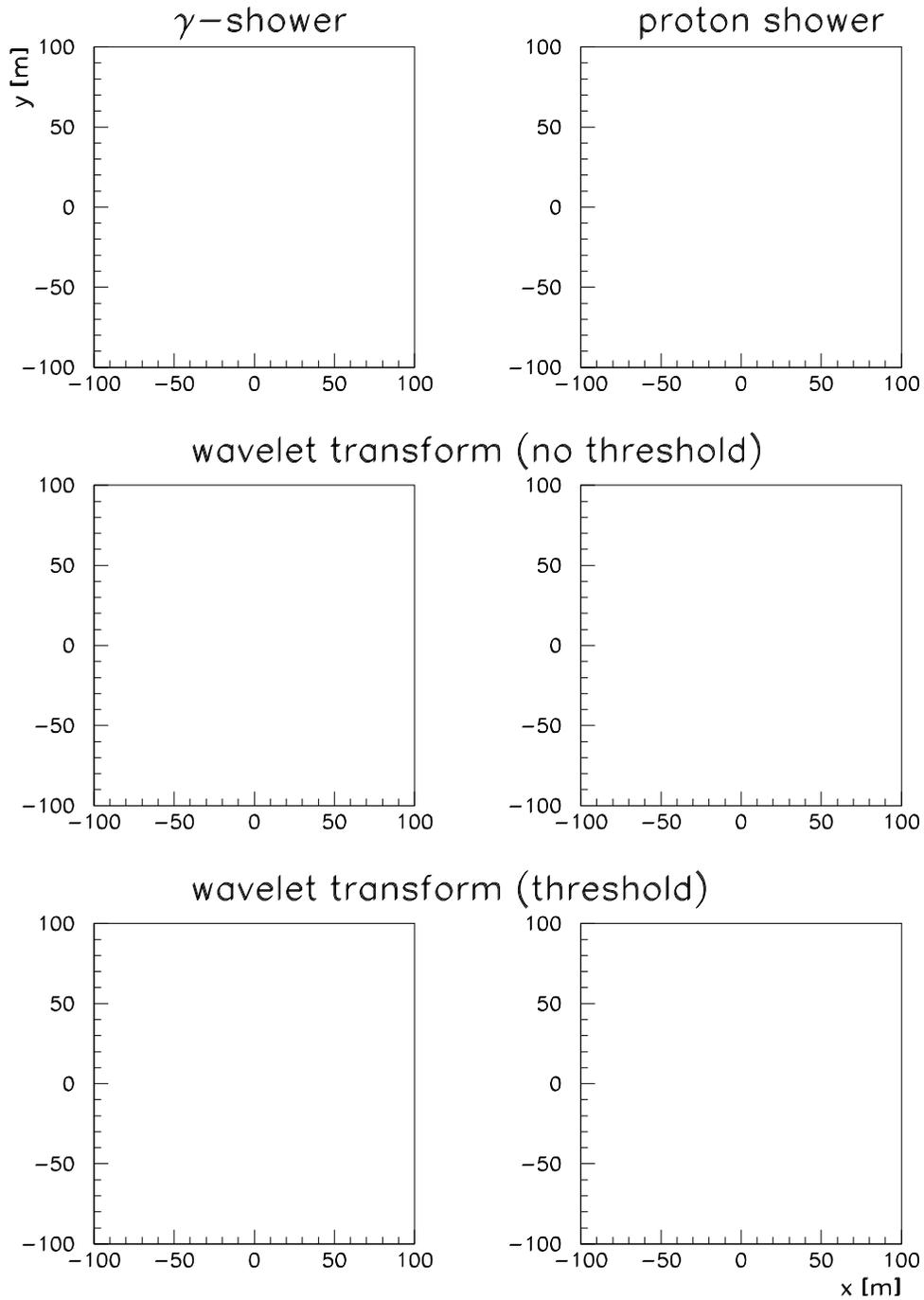,width=14.0cm}
\caption{\label{wavelet} Shower image (spatial particle 
distribution reaching ground level) for a typical 
TeV $\gamma$- and proton shower (top). Event image after 
convolution with ``Urban Sombrero'' smoothing function 
(middle), and after significance thresholding (bottom). 
(0,0) is the center of the detector.}
\end{figure}

The rate of VHE $\gamma$-ray induced showers is significantly 
smaller than those produced by
hadronic cosmic-rays\footnote{At 1 TeV the ratio of 
proton- to $\gamma$-induced showers
from the Crab Nebula is approximately 10$^4$, assuming an 
angular resolution of 0.5 degrees.}.
Therefore rejecting this hadronic background and improving 
the signal to noise ratio is crucial to
the success of any VHE $\gamma$-ray telescope. The effectiveness 
of a background rejection technique is typically expressed as a 
quality factor $Q$ defined as

\begin{equation}
Q = \frac{\epsilon_\gamma}{\sqrt{1-\epsilon_p}}
\label{equation2}
\end{equation}

where $\epsilon_\gamma$ and $\epsilon_p$ are the efficiencies 
for {\em identifying} $\gamma$-induced and 
proton-induced showers, respectively. Traditional extensive 
air-shower experiments have addressed 
$\gamma$/hadron-separation (i.\,e. background rejection) 
by identifying the penetrating particle 
component of air showers (see e.\,g. \cite{hegra_gh,casa_gh}), 
particularly muons. Although valid at 
energies exceeding 50 TeV, the number of muons detectable 
by a telescope of realistic effective area is 
small at TeV energies (see Figure\,\ref{n_muon}\,(a)). 
In addition, the $N_{\mu}$-distribution deviates 
from a Poisson distribution, implying that the fraction of 
proton showers {\em without} any muon is larger than 
$e^{-\overline{N}_{\mu}}$ (Figure\,\ref{n_muon}\,(b)). 
Relying on muon detection for an effective 
$\gamma$/hadron-separation requires efficient muon detection 
over a large area. A fine-grained absorption
calorimeter to detect muons and perform air shower calorimetry 
can, in principle, lead to an effective rejection 
factor; however, the costs associated with such a detector 
are prohibitive.

In contrast to air-shower experiments, imaging air-Cherenkov 
telescopes have achieved quality factors 
$Q>$\,7 by performing a {\em shape analysis} on the observed 
image \cite{whipple_q}. Non-uniformity of hadronic 
images arises from the development of localized regions of 
high particle density generated by small 
sub-showers. Although some of the background rejection 
capability of air-Cherenkov telescopes is a result of their 
angular resolution, rejection of hadronic events by 
identifying the differences between $\gamma$- and 
proton-induced images considerably increases source sensitivity.

Although air-Cherenkov telescopes image the shower as it 
(primarily) appears at shower maximum, these 
differences should also be evident in the particle distributions 
reaching the ground. Figure\,\ref{wavelet}\,(top) shows 
the particle distributions reaching ground level for 
typical TeV $\gamma$-ray and proton-induced
showers. This figure illustrates the key differences: 
the spatial distribution of particles in $\gamma$-ray 
showers tends to be compact and smooth, while in proton 
showers the distributions are clustered and uneven. 
Mapping the spatial distribution of shower particles 
(imaging), and identifying/quantifying shower features 
such as these should yield improved telescope sensitivity.

\subsection{$\sigma_\theta$}

Shower particles reach the ground as a thin disk of diameter 
approximately 100\,m. To first order, the disk
can be approximated as a plane defined by the arrival times 
of the leading shower front particles.
The initiating primary's direction is assumed to be a 
perpendicular to this plane. Ultimately, the accuracy with which 
the primary particle's direction can be determined is related 
to the accuracy and total number of the relative arrival 
time measurements of the shower particles, 
\begin{equation}
\sigma_{\theta} \propto \frac{\sigma_{t}}{\sqrt{\rho}}~,
\end{equation}
where $\sigma_t$ is the time resolution and $\rho$ 
is the density of independent detector elements 
sampling the shower front. The telescope must, therefore, 
be composed of elements that have fast timing $\sigma_t$
and a minimum of cross-talk since this can affect the 
shower front arrival time
determinations. Once the detector area is larger than the 
typical lateral extent of air showers, thus providing 
an optimal lever arm, the angular resolution can be further 
improved by increasing the sampling density.

To achieve ``shower limited'' resolution, individual 
detector elements should have a time response
no larger than the fluctuations inherent in shower particle 
arrival times ($\leq10$ ns, see 
Figure\,\ref{n_muon}\,(c)); on the other hand, there is no 
gain if $\sigma_t$ is significantly smaller than the 
shower front fluctuations. 

In practice fitting the shower plane is complicated by 
the fact that the shower particles undergo 
multiple scattering as they propagate to the ground leading 
to a curvature of the shower front. This scattering 
delays the particle arrival time by $\cal{O}$(ns)/100\,m, 
however the actual magnitude of curvature is a function 
of the particle's distance from the core. Determination 
of the core position, and the subsequent application of a 
{\em curvature correction} considerably improves the angular 
resolution by returning the lateral particle 
distribution to a plane which can then be reconstructed. 
Core location accuracy can be improved by increasing 
the sampling density of detector elements and the overall 
size of the detector itself.

\section{Conceptual Design}

To summarize the previous sections, an all-sky VHE telescope 
should satisfy the following design considerations:
\begin{itemize}
\item{$\sim\,100\,\%$ duty cycle ($T$)}
\item{large effective area ($A_{eff}$)}
\item{high altitude ($>$\,4000\,m)}
\item{high sampling density}
\item{fast timing}
\item{imaging capability}
\end{itemize}

In the sections that follow, we study how a pixellated 
{\em scintillator-based} large-area
detector with 100$\%$ active sampling performs as an 
all-sky monitor and survey instrument. Scintillator is 
used since it can provide excellent time resolution and has 
high sensitivity to charged particles, ultimately 
leading to improvements in angular resolution, energy threshold, 
and background rejection. To reduce detector 
cross-talk, improve timing, and enhance the imaging 
capabilities the detector should be segmented into 
optically isolated pixels . This type of detector is easier 
to construct, operate, and maintain compared to 
other large-area instruments such as water- or gas-based 
telescopes - advantageous since the high altitude 
constraint is likely to limit potential telescope sites 
to remote locations. 

\begin{figure}
\epsfig{file=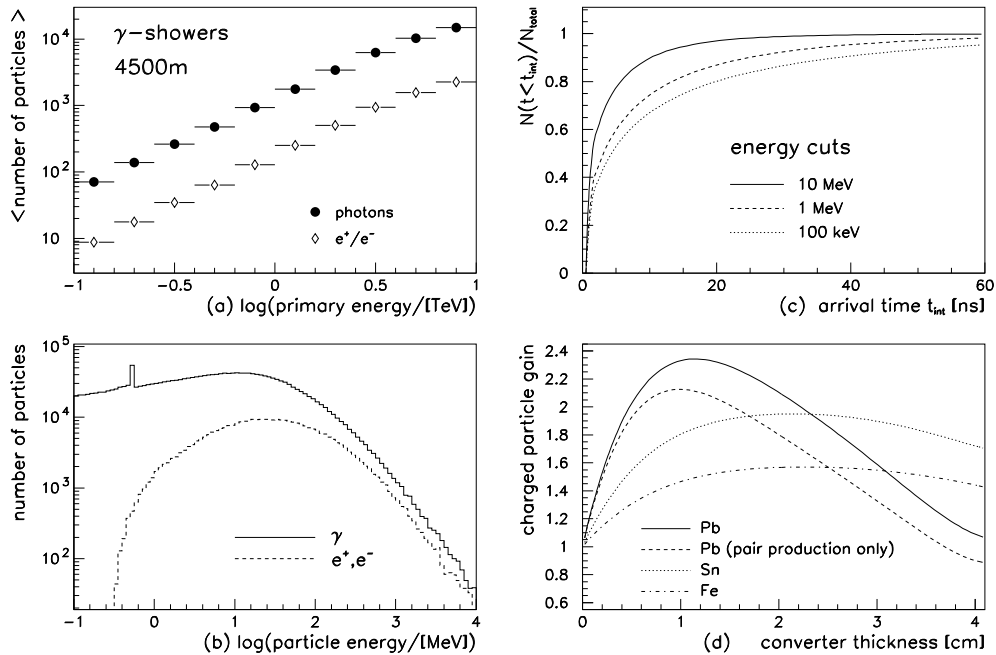,width=14.0cm}
\caption{\label{converter}
Mean number vs. primary energy (a) and energy distribution 
(b) of secondary $\gamma$'s and $e^{\pm}$ 
reaching 4500\,m observation altitude. (c) Integral 
shower particle arrival time distribution 
for particles within 40\,m distance to the core and various 
cuts on the particle energy (100\,keV, 1\,MeV, 10\,MeV).
(d) Charged particle gain as a function of the converter 
thickness for lead, tin, and iron converters.}
\end{figure}

Many of the design goals are most effectively achieved by 
maximizing the number of detected air-shower 
particles. As discussed in Section\,\ref{sec_alt}, detector 
altitude is of primary importance; however, at the energies 
of interest here only about $10\,\%$ 
(Figure\,\ref{converter}\,(a,b)) of the particles reaching 
the detector level are charged. Thus, the number of detected 
particles can be increased dramatically by improving the 
sensitivity to the $\gamma$-ray component of showers. 
A converter material (e.\,g. lead) 
on top of the scintillator converts photons into charged 
particles via Compton scattering and pair production,
and, in addition, reduces the time spread of the shower 
front by filtering low energy particles
which tend to trail the prompt shower front 
(Figure\,\ref{converter}\,(c)) and thus 
deteriorate the angular resolution.
Figure\,\ref{converter}\,(d) shows the charged particle 
gain expected as a function of the converter thickness 
for lead, tin, and iron converters. 
The maximum gain is for a lead converter at $\sim$\,2 
radiation lengths 
($1\,{\mathrm r.\,l.}=0.56\,{\mathrm cm}$), but the gain 
function is rather steep below 1\,r.\,l. and flattens 
above. Because of the spectrum of secondary $\gamma$-rays 
reaching the detector pair production is the most 
dominant process contributing to the charged particle gain 
(Figure\,\ref{converter}\,(d)).
 
Techniques for reading out the light produced in 
scintillator-based detector elements have progressed
in recent years with the development of large-area sampling 
calorimeters. Of particular interest
is the work by the CDF collaboration on scintillating 
tiles \cite{bodek}; this technique utilizes fibers, 
doped with a wavelength shifter and embedded directly in 
the scintillator, to absorb the scintillation light 
and subsequently re-emit it at a longer wavelength. 
This re-emitted light is then coupled to photomultiplier 
tubes either directly or using a separate clear fiber-optic 
light guide. This highly efficient configuration 
is ideal for detecting minimum ionizing particles (MIPs), 
and produces 4 photoelectrons/MIP on average 
in a 5\,mm-thick scintillator tile.

Using an array of tile/fiber detector elements one can now 
consider a large-area detector that counts
particles and is $\sim\,100\,\%$ active, and it is this 
paradigm that we discuss in more detail in the 
following sections. It should be noted that a 
scintillator-based air-shower detector is not a new idea; 
however, the use of the efficient tile/fiber configuration 
in a detector whose physical area is fully active 
pushes the traditional concept of an air-shower array 
to the extreme.

\subsection{Air Shower and Detector Simulation}

The backbone of a conceptual design study is the simulation 
code. Here the complete simulation of
the detector response to air showers is done in two steps: 
1) initial interaction of the primary particle 
(both $\gamma$-ray and proton primaries) with the atmosphere 
and the subsequent development
of the air shower, and 2) detector response to air-shower 
particles reaching the detector level.

The CORSIKA \cite{corsika} air shower simulation code, 
developed by the KASCADE \cite{kascade} group, 
provides a sophisticated simulation of the shower development 
in the Earth's atmosphere. In CORSIKA, 
electromagnetic interactions are simulated using the 
EGS\,4 \cite{egs} code. For hadronic interactions, 
several options are available. A detailed study of the hadronic 
part and comparisons to existing data has 
been carried out by the CORSIKA group and is documented in 
\cite{corsika}. For the simulations discussed here, 
we use the VENUS \cite{venus} code for high energy hadronic 
interactions and GHEISHA \cite{gheisha} to treat low energy 
($\le\,80\,{\mathrm GeV}$) hadronic interactions.

The simulation of the detector itself is based on the 
GEANT \cite{geant321} package. The light yield of 
0.5\,cm tile/fiber assemblies has been studied in detail 
in \cite{bodek} and \cite{barbaro}, and we adopt 
an average light yield of 4 photoelectrons per minimum 
ionizing particle. This includes attenuation losses 
in the optical fibers and the efficiency of the photomultiplier. 
Simulation parameters are summarized in 
Table\,\ref{tab_sim}. It should be noted that wavelength 
dependencies of the fiber attenuation length and 
of the photomultiplier quantum efficiency have not been included.

\begin{table}
\caption{\label{tab_sim} Basic parameters of the shower and 
detector simulation.}
\vskip 0.5cm
\small
\centerline{
\begin{tabular}{|l|l|}
\hline
zenith angle range  & $0^{\mathrm o}\le\theta\le45^{\mathrm o}$ \\
lower kinetic energy cuts & 0.1\,MeV ($e^{\pm},\,\gamma$)    \\
                          & 0.1\,GeV ($\mu^{\pm}$)           \\
                          & 0.3\,GeV (hadrons)               \\
scintillator thickness    & 0.5\,cm                          \\
lead converter thickness  & 0.5\,cm                          \\
PMT transit time spread   & 1\,ns (FWHM)                     \\
average light yield/MIP   & 4 photoelectrons                 \\
\hline
\end{tabular}}
\end{table}
\normalsize

As shown in Section\,\ref{sec_alt} only a detector at an 
altitude above 4000\,m can be expected to give the desired 
performance; however, we study the effect of three 
detector altitudes, 2500\,m 
($764.3\,{\mathrm g}\,{\mathrm cm}^{-2}$), 3500\,m 
($673.3\,{\mathrm g}\,{\mathrm cm}^{-2}$), and 
4500\,m ($591.0\,{\mathrm g}\,{\mathrm cm}^{-2}$), for the 
purpose of completeness. These altitudes span 
the range of both existing and planned all-sky VHE telescopes 
such as the Milagro detector \cite{milagro} near Los Alamos, 
New Mexico (2630\,m a.s.l.), and the ARGO-YBJ \cite{argo} detector 
proposed for the Yanbajing Cosmic Ray Laboratory in Tibet 
(4300\,m a.s.l.).

\section{Detector Performance}

In the remainder of this paper we study the expected performance 
of a detector based on the conceptual design 
discussed above. Although the canonical design is a detector 
with a geometric area $150\times 150\,\mathrm{m}^{2}$, a detector 
design incorporating a $200\times 200\,\mathrm{m}^{2}$ area has
also been analyzed in order to understand how telescope 
performance scales with area. Pixellation is achieved
by covering the physical area of the detector with a mosaic 
of 5\,mm thick scintillator tiles each covering
an area of $1\times 1\,\mathrm{m}^{2}$.

\subsection{Energy Threshold} 

\begin{figure}
\epsfig{file=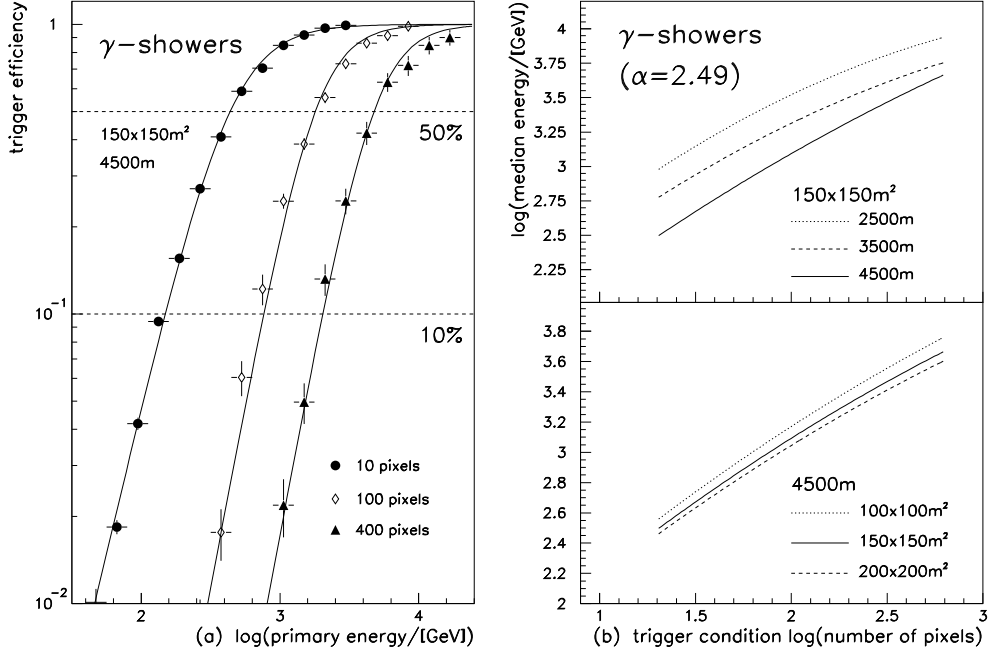,width=14.0cm}
\caption{\label{energy}
(a) Trigger efficiency as a function of the primary 
particles' energy for three trigger conditions 
(10, 40, 400 pixels).
(b) Median energy of detected $\gamma$-showers as a function 
of the trigger condition (number
of pixels) for three detector altitudes (top) and as a 
function of the detector size for 
a fixed altitude (4500\,m) (bottom).}
\end{figure}

The energy threshold of air-shower detectors is not well-defined. 
The trigger probability for a shower 
induced by a primary of fixed energy is not a step-function but 
instead rises rather slowly due 
to fluctuations in the first interaction height, shower 
development, core positions, and incident angles. 
Figure\,\ref{energy}\,(a) shows the trigger probability as a 
function of the primary $\gamma$-ray energy 
for three trigger conditions. 

Typically, the primary energy where the trigger probability 
reaches either $10\,\%$ or $50\,\%$ is defined as the 
energy threshold (see Figure\,\ref{energy}\,(a)).
A large fraction of air showers that fulfill the trigger 
condition will have lower energies since 
VHE source spectra appear to be power-laws, $E^{-\alpha}$. A 
more meaningful indication of energy threshold 
then is the {\it median} energy $E_{med}$; however, this 
measure depends on the spectral index of the source. 
For a source with spectral index $\alpha=2.49$ (Crab), 
Figure\,\ref{energy}\,(b, top) shows $E_{med}$ as a 
function of detector altitude (for fixed detector size). 
The median energy increases as altitude decreases 
since the number of particles reaching the detector level 
is reduced at lower altitudes. For a detector at 
4500\,m a.s.l., $E_{med}$ is about 500\,GeV after imposing 
a 40 pixel trigger criterion.

$E_{med}$ is not a strong function of the detector size 
(Figure\,\ref{energy}\,(b, bottom)). It is 
also noteworthy that a larger pixel size of 
$2\times 2\,\mathrm{m}^2$ instead of $1\times 1\,\mathrm{m}^2$
only slightly increases $E_{med}$; due to the lateral extent 
of air showers and the large average distances 
between particles, nearly $95\,\%$ of all showers with more 
than 50 $1\,\mathrm{m}^2$-pixels also have more 
than 50 $4\,\mathrm{m}^2$-pixels.

\subsection{Background Rejection and Core Location}
\label{sec_bg}

Due to its pixellation and $100\%$ active area the telescope 
described here can provide true images of 
the spatial distribution of secondary particles reaching the 
detector. Image analysis can take many forms; 
the method of wavelet transforms\,\cite{kaiser} is well 
suited for identifying and extracting localized 
image features. To identify localized high-density regions 
of particles an image analysis technique that utilizes digital 
filters is used; the procedure is briefly summarized below 
while details are given in \cite{miller}.

Proton- and $\gamma$-induced showers can be identified by 
counting the number of ``hot spots'', or peaks, in a 
shower image in an automated, unbiased way. These peaks are 
due to small sub-showers created by secondary particles and 
are more prevalent in hadronic showers than in $\gamma$-induced 
showers. To begin, the shower image (i.\,e. the 
spatial distribution of detected secondary particles) is 
convolved with a function that smooths the image over a 
predefined region or length scale 
(see Figure\,\ref{wavelet}\,(middle)). 
The smoothing function used in this 
analysis is the so-called ``Urban Sombrero''\footnote{Also 
known as the ``Mexican Hat'' function.} function:
\begin{equation}
g\left(\frac{r}{a}\right) 
= \left(2 - \frac{r^2}{a^2}\right) e^{-\frac{r^2}{2\,a^2}}
\end{equation}
where $r$ is the radial distance between the origin of 
the region being smoothed and an image pixel, and $a$ is the 
length scale over which the image is to be smoothed. 
This function is well suited for this analysis since it is a 
localized function having zero mean; therefore, image 
features analyzed at multiple scales $a$ will maintain their 
location in image space. A peak's maximum amplitude is found 
on length scale $a$ corresponding to the actual 
spatial extent of the ``hot spot.''
 
\begin{figure}
\epsfig{file=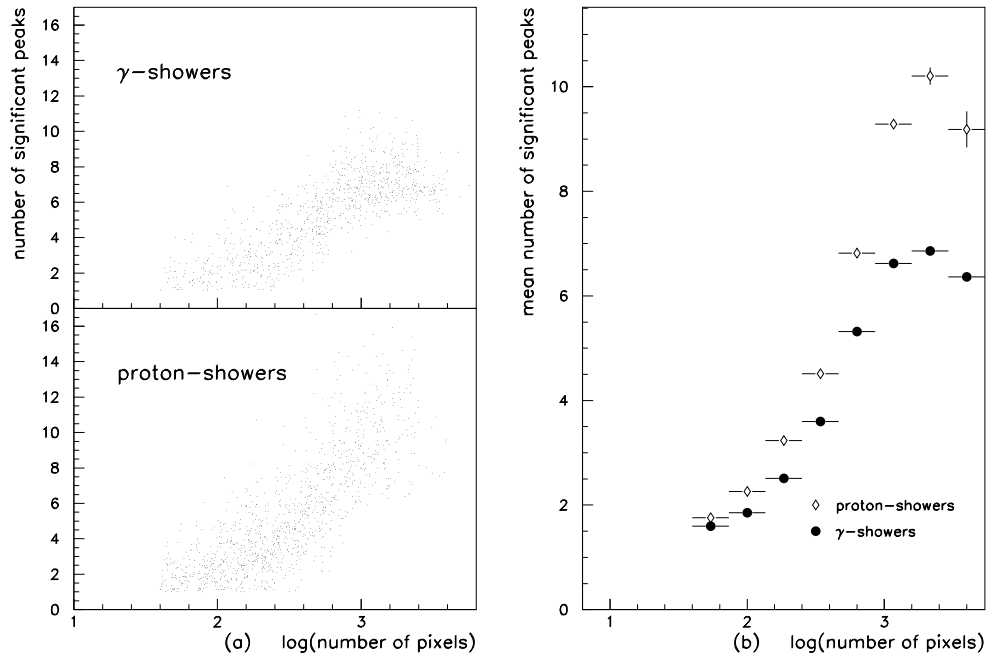,width=14.0cm}
\caption{\label{peaks} 
(a) Number of significant peaks for $\gamma-$ (top) and 
proton showers (bottom) as a function of pixel multiplicity. 
A random probability of $<\,6.3\times10^{-5}$ is used to 
define a significant peak.
(b) Mean number of significant peaks for $\gamma-$ and 
proton showers as a function of pixel multiplicity.}
\end{figure}

Many peaks exist in these images; the key is to tag 
statistically significant peaks. To do this the probability 
distribution of peak amplitudes must be derived from a random 
distribution of pixels. Using $2\times10^4$ events,
each with a random spatial particle distribution, 
the probability of observing a given amplitude is computed. This
is done for events with different pixel multiplicities and 
for different scale sizes. Results using only a 
single scale size of 8\,m are presented here. This scale size 
represents the optimum for the ensemble of showers; 
scale size dependence as a function of pixel multiplicity is 
studied in \cite{miller}. In order to identify 
statistically significant peaks a threshold is 
applied to the smoothed image. Peaks are eliminated if the 
amplitude was more probable than $6.3\times10^{-5}$
corresponding to a significance of $<4\,\sigma$; the value 
of the threshold is chosen to maximize the background 
rejection. Figure\,\ref{wavelet}\,(bottom) shows the result 
of thresholding. After applying thresholding 
the number of significant peaks is counted; if the number of 
peaks exceeds the mean number of peaks expected from 
$\gamma$-induced showers then the event is tagged as 
a ``proton-like'' shower and rejected. The number of 
expected peaks is energy dependent starting at 1 peak 
(i.\,e. the shower core), on average, for $\gamma$-showers 
with less than 100 pixels and increasing with pixel 
multiplicity; proton-induced showers show a similar 
behavior except 
that the number of peaks rises faster with the number of 
pixels (see Figure\,\ref{peaks}).

Additional background rejection may be possible by using 
information such as the spatial distribution of peaks and 
the actual shape of individual peak regions. This is 
currently being investigated. The ability to map 
and analyze the spatial distributions of air shower particles 
implies that, in conjunction with analysis
techniques such as the one described here, the large cosmic-ray 
induced backgrounds can be suppressed thereby 
improving the sensitivity of a ground-based air-shower array; 
quantitative results on the use of this image analysis technique 
are described below and summarized in Figure\,\ref{quality}.

\begin{figure}
\epsfig{file=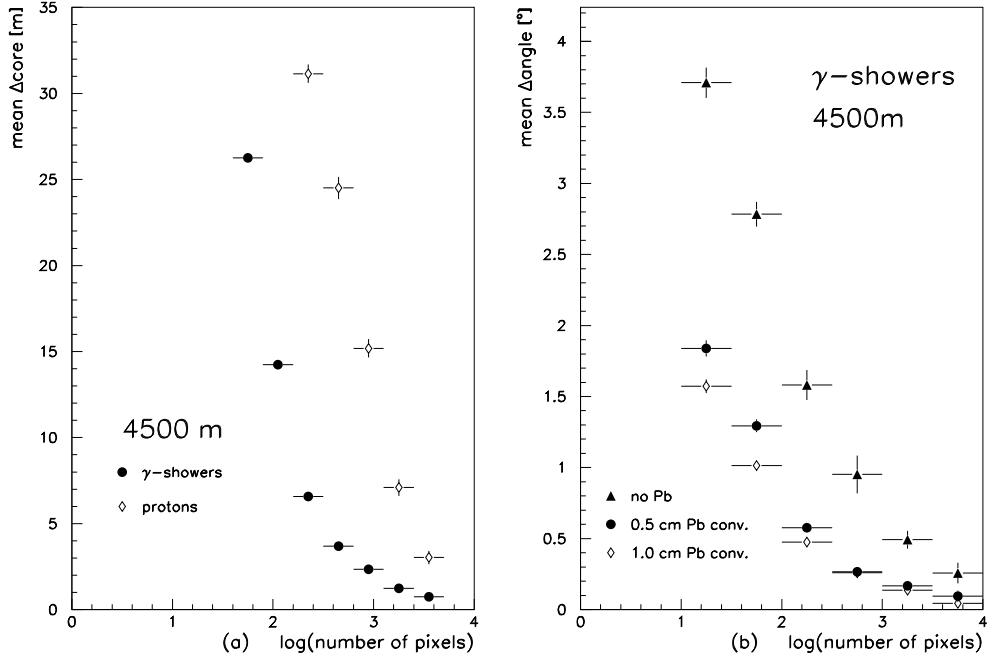,width=14.0cm}
\caption{\label{angle_res} 
(a) Mean distance between reconstructed and true shower 
core location.
(b) Mean angle between reconstructed and true shower direction for
a detector without lead converter and with 0.5\,cm and 
1.0\,cm lead.}
\end{figure}

Image analysis can also be used to identify and locate the 
shower core. Here the core is identified as the
peak with the largest amplitude; this is reasonable assuming 
that typically the core represents a relatively
large region of high particle density. 
Figure\,\ref{angle_res}\,(a) shows the accuracy of the core fit; 
other methods are less accurate and more susceptible 
to detector edge effects
and local particle density fluctuations.

The core location is used to correct for the curvature of the 
shower front and to veto events with cores outside the active 
detector area. Rejecting ``external'' events is beneficial
as both angular resolution and background rejection 
capability are worse for events with cores off the detector.
We define the outer 10\,m of the detector as a veto ring 
and restrict the analysis to events with
fitted cores inside the remaining fiducial area 
($130\times 130\,\mathrm{m}^{2}$ or 
$180\times 180\,\mathrm{m}^{2}$ for the 
$200\times 200\,\mathrm{m}^{2}$ design). This cut 
identifies and keeps $94\,\%$ of the $\gamma$-showers 
with {\em true} cores within the fiducial area while 
vetoing $64\,\%$ of events with cores outside. It is 
important to note that the non-vetoed events are generally 
of higher quality (better angular resolution, 
improved $\gamma$/hadron-separation). In addition, $R_{\gamma}$ is 
smaller for external events than for internal ones 
due to the larger lateral spread of particles in proton 
showers; thus the veto cut actually improves overall 
sensitivity even though total effective area is decreased. 
 
\subsection{Angular Resolution}

The shower direction is reconstructed using an iterative 
procedure that fits a plane to the arrival times of the pixels and 
minimizes $\chi^{2}$.
Before fitting, the reconstructed core position 
(see Section\,\ref{sec_bg}) is used
to apply a curvature correction to the shower front.
In the fit, the pixels are weighted with $w(p)=1/\sigma^{2}(p)$, 
where $\sigma^{2}(p)$ is the RMS of the time residuals 
$t_{pixel}-t_{fit}$ for pixels with $p$ photoelectrons.
$t_{fit}$ is the expected time according to the previous iteration.

In order to minimize the effect of large time fluctuations 
in the shower particle arrival times, we reject pixels with times 
$t_{pixel}$ with 
$\left|t_{pixel}-t_{fit}\right|\ge\,10\,\mathrm{ns}$.
In addition, only pixels within 80\,m distance to the 
shower core are included in the fit.

Figure\,\ref{angle_res}\,(b) shows the mean difference
between the fitted and the true shower direction as a 
function of the number of pixels for a 
detector with and without 0.5\,cm and 1.0\,cm of lead, 
again indicating the benefits of the 
converter. The angular resolution does not improve considerably 
when the converter thickness is increased from 0.5\,cm 
to 1.0\,cm, thus 0.5\,cm is a reasonable
compromise considering the tradeoff between cost and performance.

\subsection{Sensitivity}

\begin{figure}
\epsfig{file=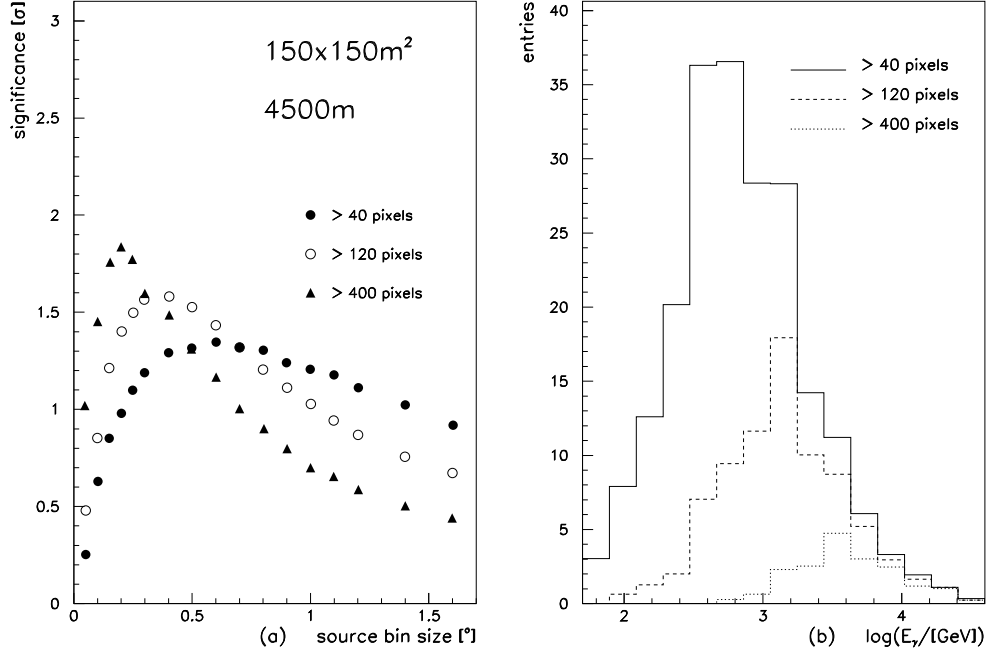,width=14.0cm}
\caption{\label{sourcebin}
(a) Significance for a one day observation of a 
Crab-like source for three trigger conditions 
(40, 120, 400 pixels) as a function of the source bin size.
(b) Energy distribution of the 
detected $\gamma$-showers for the three trigger conditions. 
The source location is
$\left|\delta-\lambda\right|\simeq\,5^{\mathrm{o}}$,
where $\delta$ is the source declination and $\lambda$ 
is the latitude of the detector site.}
\end{figure}

\begin{figure}
\epsfig{file=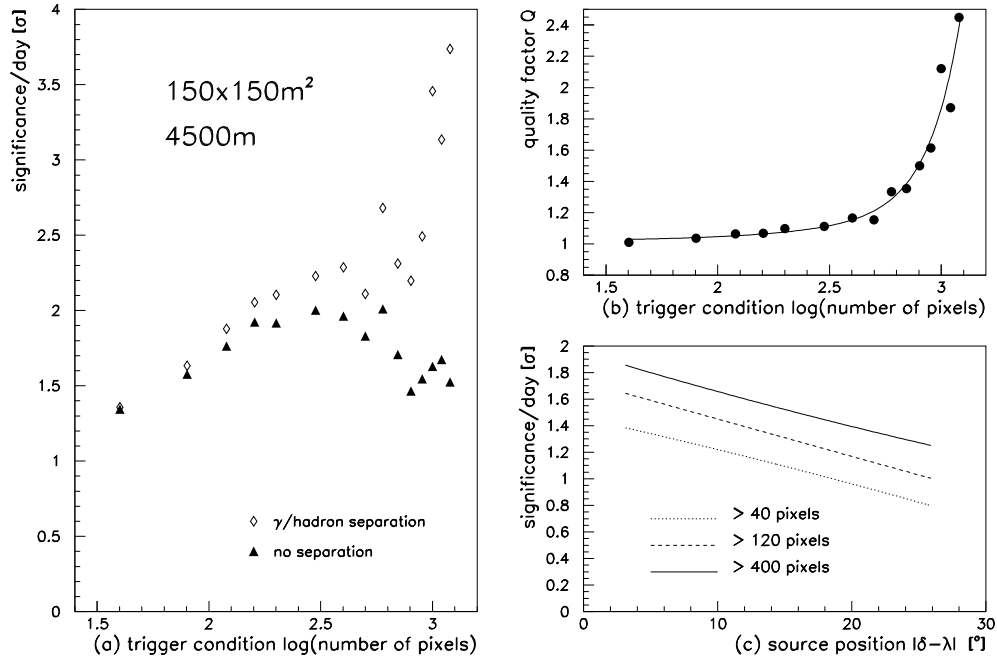,width=14.0cm}
\caption{\label{quality}
(a) Significance/day for a Crab-like source as a function 
of the trigger condition with and without 
$\gamma$/hadron-separation for the $150\times 150\,\mathrm{m}^{2}$ 
prototype.  (b) Quality factor as a function of the 
trigger condition. (c) Sensitivity as a function of the source 
position $\left|\delta-\lambda\right|$.}
\end{figure}

The ultimate characteristic of a detector is its sensitivity 
to a known standard candle.
In this section, the methods described so far are 
combined to estimate the overall
point source sensitivity of a pixellated scintillation detector.
As indicated in Equation\,\ref{equation1}, the sensitivity 
of an air shower array depends on its 
angular resolution $\sigma_{\theta}$, its effective area 
$A_{eff}$, the trigger probabilities 
for source and background showers, and the quality factor of the 
$\gamma$/hadron-separation. However, as most of the parameters
are functions of the primary energy, the sensitivity depends on the
spectrum of the cosmic ray background and the spectrum of the 
source itself. The significance $S$ therefore has to be 
calculated using
\begin{equation}
S=\frac{\int A^{eff}_{\gamma}(E)~\epsilon_{\gamma}(E)
~J_{\gamma}(E)~{\mathrm d}E~~f_{\gamma}~T}
       {\sqrt{\int A^{eff}_{p}(E)~(1\,-\,\epsilon_{p}(E))
~J_{p}(E)~{\mathrm d}E~~\Delta\Omega~T}}
\label{equation4}
\end{equation}
where $J_{\gamma}$ and $J_{p}$ are the photon and proton 
energy spectrum, and
$f_{\gamma}$ is the fraction of $\gamma$-showers 
fitted within the solid angle bin
$\Delta\Omega=2\,\pi\,(1-\mathrm{cos}\,\theta)$. 
Other parameters have their standard meaning.

The Crab Nebula is commonly treated as a standard candle 
in $\gamma$-ray astronomy; this allows 
the sensitivity of different telescopes to be compared. 
The differential spectrum of the Crab 
at TeV energies has been measured by the 
Whipple collaboration \cite{whipple_crab}:
\begin{equation}
J_{\gamma}(E) = (3.20\,\pm0.17\,\pm0.6)\times10^{-7}\,
E_{\mathrm{TeV}}^{-2.49\,\pm0.06\,\pm0.04}\,
\mathrm{m}^{-2}\,\mathrm{s}^{-1}\,\mathrm{TeV}^{-1}.
\label{crab_rate}
\end{equation}
The sensitivity of the detector to a Crab-like source 
can be estimated using Equation\,\ref{equation4} and 
the differential proton background flux measured by the 
JACEE balloon experiment \cite{jacee}:
\begin{equation}
\frac{dJ_{p}(E)}{d\Omega}=(1.11^{+0.08}_{-0.06})\times10^{-1}\,
E_{\mathrm{TeV}}^{-2.80\,\pm0.04}\,
\mathrm{m}^{-2}\,\mathrm{sr}^{-1}\,\mathrm{s}^{-1}\,
\mathrm{TeV}^{-1}.
\label{bg_rate}
\end{equation}

Calculating $S$ using Equation\,\ref{equation4} is not 
straightforward since $\epsilon_{\gamma}$
is not a constant, but rather a function of the event size 
and core position and, therefore, angular reconstruction accuracy. 
This equation can be solved, however, by a Monte Carlo 
approach: Using the Crab and cosmic-ray proton spectral 
indices, a pool of simulated $\gamma$- and proton showers 
($\cal{O}$($10^{6}$) events of each particle type at
each altitude) is generated with energies from 50\,GeV 
to 30\,TeV and with zenith angles 
$0^{\mathrm o}\ge\theta\ge45^{\mathrm o}$. 
A full Julian day source transit can be simulated for a given 
source bin size and declination 
by randomly choosing $\gamma$-ray and proton-induced showers 
from the simulated shower pools at rates
given by Equations\,\ref{crab_rate} and~\ref{bg_rate}. The only 
constraint imposed on the events is that 
they have the same zenith angle as the source bin at the 
given time. The showers are then fully reconstructed, 
trigger and core location veto cuts, as well as 
$\gamma$/hadron-separation cuts, are applied; $\gamma$-showers 
are required to fall into the source bin. This procedure 
produces distributions of pixel multiplicity, core 
position, etc. reflecting instrumental resolutions and responses. 

Because the angular resolution varies with the number of 
pixels, the optimal source bin size also varies. 
Figure\,\ref{sourcebin}\,(a) shows the significance for a full-day
observation of a Crab-like source with 
$\left|\delta-\lambda\right|\simeq\,5^{\mathrm{o}}$,
where $\delta$ is the source declination and $\lambda$ 
is the latitude of the detector site,
as a function of the source bin size for three trigger conditions. 
As the number of pixels increases, the optimal source bin size 
decreases from $0.6^{\mathrm{o}}$ (40 pixels) to 
$0.2^{\mathrm{o}}$ (400 pixels). 
Figure\,\ref{sourcebin}\,(b) shows 
how the energy distribution of detected $\gamma$-showers 
changes with trigger condition. The 
median energy for a 40 pixel trigger is 600\,GeV, 
with a substantial fraction of showers having energies
below 200\,GeV. For a 120 pixel trigger, $E_{med}$ is 1\,TeV.

For a 1 day Crab-like source transit, Figure\,\ref{quality}\,(a) 
shows how the significance varies as a 
function of the trigger condition with and without 
$\gamma$/hadron-separation. If no $\gamma$/hadron-separation
is applied the sensitivity increases and then falls above 
500 pixels because of the finite size of the detector. 
However, as shown in Section\,\ref{sec_bg}, above 500 pixels 
the quality factor of the 
$\gamma$/hadron-separation counterbalances the loss of area. 
Figure\,\ref{quality}\,(b) shows the quality factor 
derived solely from the ratio of source significance with 
and without separation. As expected, $Q$ increases 
dramatically with pixel number, leading to significances 
well above $3\,\sigma$ per day for energies above 
several TeV.
The sensitivity also depends on the declination of the source. 
Results quoted so far refer to sources
$\left|\delta-\lambda\right|\simeq\,5^{\mathrm{o}}$. 
Figure\,\ref{quality}\,(c) shows how the
expected significance per source day transit changes with 
the source declination $\delta$.

\begin{table}[ht]
\caption{\label{table2} Significances $S$ for a 1 day 
observation of a Crab-like source with 
$\left|\delta-\lambda\right|\simeq\,5^{\mathrm{o}}$
for different altitudes and trigger conditions. $E_{med}$ is the
median energy of the detected source particles, values 
in parentheses denote significances after 
$\gamma$/hadron separation.}
\vskip 0.5cm
\small
\centerline{
\begin{tabular}{|c|c|| c c | c || c c | c |}
\hline
   &    & \multicolumn{3}{|c||}{$150\times 150\,\mathrm{m}^{2}$} 
        & \multicolumn{3}{|c|}{$200\times 200\,\mathrm{m}^{2}$} \\
\hline
altitude & trigger & \multicolumn{2}{|c|}{$S \left[\sigma\right]$} 
& log($E_{med}^{\mathrm{GeV}})$
         & \multicolumn{2}{|c|}{$S \left[\sigma\right]$} 
& log($E_{med}^{\mathrm{GeV}})$ \\
\hline
\hline
4500\,m & 40   & 1.3 & (1.3) & 2.8 & 1.8 & (1.8) & 2.8 \\
        & 1000 & 1.6 & (2.5) & 3.8 & 1.9 & (2.7) & 3.8 \\
\hline
3500\,m & 40   & 0.9 & (0.9) & 3.1 & 1.2 & (1.2) & 3.0 \\
        & 1000 & 0.8 & (1.2) & 4.1 & 1.3 & (1.7) & 4.0 \\
\hline 
2500\,m & 40   & 0.4 & (0.4) & 3.3 & 0.6 & (0.6) & 3.2 \\
        & 1000 & 0.5 & (0.6) & 4.3 & 0.8 & (1.1) & 4.2 \\
\hline 
\end{tabular}}
\end{table}
\normalsize

\begin{table}[ht]
\caption{\label{table3} Expected rates [kHz] for a 
$150\times 150\,\mathrm{m}^2$ detector at 
different altitudes. Cores are randomly distributed 
over $300\times 300\,\mathrm{m}^{2}$ and no core
veto cut is applied.}
\vskip 0.5cm
\small
\centerline{
\begin{tabular}{|l||c|c|c|}
\hline
trigger & 4500\,m & 3500\,m & 2500\,m \\
\hline
\hline
10   & 34.5  & 18.9  & 10.5  \\
40   &  6.7  &  3.6  &  2.1  \\
100  &  1.8  &  1.0  &  0.6  \\
400  &  0.2  &  0.2  &  0.1  \\
1000 &  0.05 &  0.04 &  0.02 \\
\hline
\end{tabular}}
\end{table}
\normalsize

Table\,\ref{table2} summarizes the dependence of the 
detector performance on the size and the altitude
of the detector. Significance scales with $\sqrt{A_{eff}}$ 
as expected from Equation\,\ref{equation1}. Detector 
altitude, however, is more critical. Although at 2500\,m 
$10\,\sigma$ detections of a steady Crab-like source 
per year are possible, it is only at 4000\,m altitudes where 
the sensitivity is sufficient to detect 
statistically significant daily variations of source emissions.

It is noteworthy that for the canonical design at 4500\,m, 
a trigger condition as low as 10 pixels still produces 
$1.2\,\sigma$ per day at median energies of about 280\,GeV 
corresponding to an event rate of 34.5\,kHz.
Predicted event rates for different trigger conditions at 
various altitudes are summarized in Table\,\ref{table3}.
Sustained event rates below $\sim$\,10\,kHz are achievable 
with off-the-shelf data acquisition electronics; higher
rates may also be possible. The event rates estimated here 
are relatively low compared to the rate of 
single cosmic-ray muons; because of the optical isolation and 
low-cross talk of individual detector elements 
single muons are unlikely to trigger the detector even with 
a low pixel multiplicity trigger condition.

\section{Conclusion}

\begin{figure}
\epsfig{file=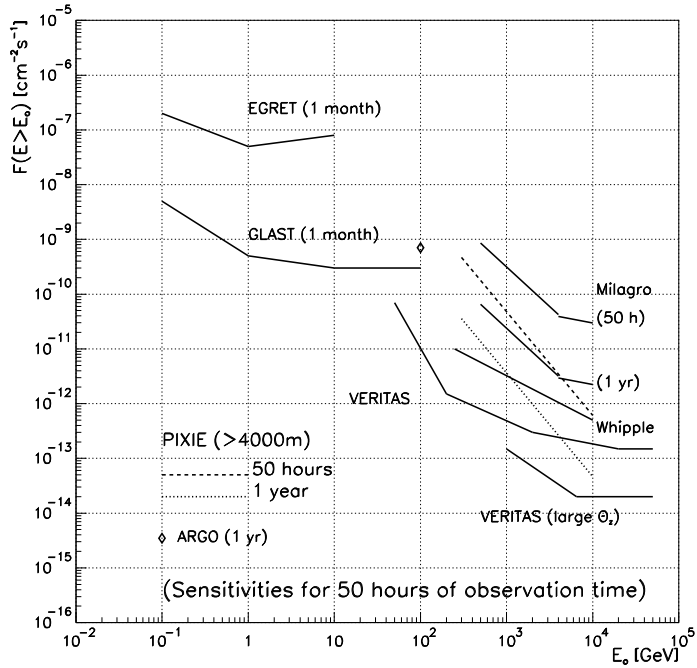,width=14.0cm}
\caption{\label{sensi_comp} Predicted sensitivity of some 
proposed and operational
ground-based telescopes. The dashed and dotted lines show 
the predicted sensitivity of the telescope described 
here (PIXIE) at an altitude $>4000$\,m. The numbers are based 
on a $5\,\sigma$ detection for the given exposure 
on a single source. EGRET and GLAST sensitivities are for 
1 month of all-sky survey. The ARGO sensitivity is 
taken from \cite{argo}, all others from \cite{glast_proposal}. 
Information required to extrapolate
the ARGO sensitivity to higher energies is not given 
in \cite{argo}.}
\end{figure}

To fully develop the field of VHE $\gamma$-ray astronomy 
a new instrument is required - one capable of 
continuously monitoring the sky for VHE transient emission 
and performing a sensitive systematic survey for 
steady sources. To achieve these goals such an instrument 
must have a wide field of view, $\sim\,100\,\%$ duty 
cycle, a low energy threshold, and background rejection 
capabilities. Combining these features we have shown that a 
detector composed of individual scintillator-based pixels 
and 100$\%$ active area provides high sensitivity 
at energies from 100\,GeV to beyond 10\,TeV. Detailed 
simulations indicate that a source with the intensity 
of the Crab Nebula would be observed with an energy dependent 
significance exceeding $\sim\,3\,\sigma$/day. 
AGN flares, or other transient phenomena, could be detected 
on timescales $\ll$1 day depending on their 
intensity - providing a true VHE transient all-sky monitor. 
A conservative estimate of the 
sensitivity of a detector like the one described here 
(the PIXIE telescope) is shown in Figure\,\ref{sensi_comp} 
compared to current and future ground- and space-based 
experiments. The plot shows the sensitivities for both 50 
hour and 1 year source exposures, relevant for transient and 
quiescent sources, respectively. A detector based on
the PIXIE design improves upon first-generation detector 
concepts, such as Milagro, in two principal ways: fast 
timing and spatial mapping of air shower particles. 
The sensitivity of a sky map produced by this detector in 
1 year reaches the flux sensitivity of current air-Cherenkov 
telescopes (for a 50 hour exposure), making the detection of 
AGN in their quiescent states possible.

The cost for a detector based on the conceptual design 
outlined is estimated conservatively at between 
\$\,500 and \$\,1000 per pixel; the cost of scintillator
is the dominant factor. A proposal to perform a detailed 
detector design study (evaluation of detector 
materials, data acquisition prototyping, and investigation 
of construction techniques) leading to a final 
design is currently pending. 

Although unlikely to surpass the sensitivity of 
air-Cherenkov telescopes for detailed single source 
observations, a sensitive all-sky VHE telescope could 
continuously monitor the observable sky at VHE energies. 
In summary, non-optical ground-based VHE astronomy {\em is} 
viable, and the development of an all-sky VHE telescope with 
sensitivity approaching that of the existing narrow field of 
view air-Cherenkov telescopes, will contribute to 
the continuing evolution of VHE astronomy.

\begin{ack}
We thank the authors of CORSIKA for providing us with the 
simulation code; we also acknowledge 
D.G. Coyne, C.M. Hoffman, J.M. Ryan, and D.A. Williams 
for their useful comments. This research is 
supported in part by the U.S. Department of Energy Office 
of High Energy Physics, the U.S. Department of Energy 
Office of Nuclear Physics, the University of California (RSM), 
and the National Science Foundation (SW).
\end{ack}

\end{document}